\documentclass[11pt]{article}

\usepackage{graphicx,epsfig,color,amssymb,amsmath,amsfonts,amscd}
\usepackage{cite,booktabs}
\usepackage{makeidx}
\usepackage{fullpage}

\newcommand{\be}{\begin{eqnarray}}
\newcommand{\ee}{\end{eqnarray}}

\usepackage[]{caption}
\def\mc{\mathcal}
\def\be{\begin{equation}}
\def\ee{\end{equation}}
\def\mpl{m_{Pl}}

\def\D{\mc D}
\def\-g{\sqrt{-g}}
\def\i{\textit}
\def\b{\textbf}
\renewcommand\tilde{\widetilde}
\renewcommand\rho{\varrho}

\newcommand\rif[1]{(\ref{#1})}

\begin{document}


\title{Curvature Oscillations in Modified Gravity and High Energy Cosmic Rays
}
\author{E.V. Arbuzova$^{a,b}$, A.D. Dolgov$^{a,c,d,e}$, and L. Reverberi$^{d,e}$}
\maketitle
\begin{flushleft}
$^a$Novosibirsk State University, Novosibirsk, 630090, Russia;\\  
$^{b}$Department of Higher Mathematics, University "Dubna", 141980 Dubna, Russia;\\
$^{c}$ ITEP
Bol. Cheremushkinsaya ul., 25, 113259 Moscow, Russia;\\ 
$^{d}$Dipartimento di
Fisica e Scienze della Terra, Universit\`a degli Studi di Ferrara,\\
Polo Scientifico e Tecnologico $-$ Edificio C, via Saragat 1, 44122 Ferrara, Italy;\\ 
$^{e}$Istituto Nazionale di Fisica Nucleare, Sezione di Ferrara,\\
Polo Scientifico e Tecnologico $-$ Edificio C, via Saragat 1, 44122 Ferrara, Italy.\\
\end{flushleft}

\begin{center}
\centering
\begin{table}[h]
\begin{tabular}{r l}
 \b{e-mail:} & arbuzova@uni-dubna.ru\\

& dolgov@fe.infn.it\\
& reverberi@fe.infn.it
\end{tabular}
\end{table}
\end{center}

%

\begin{abstract}
It is shown that F(R)-modified gravitational theories lead to curvature oscillations in astrophysical systems 
with rising energy density. The frequency and the amplitude of such oscillations could be very high and 
would lead to noticeable production of energetic cosmic ray particles. 
\end{abstract}


There are two popular phenomenological explanations of the observed accelerated universe expansion. It is 
either prescribed to existence of the so called dark energy (DE) with the equation of state:
\be
P\approx - \rho,
\label{P-rho}
\ee
where $P$ and $\rho$ are respectively pressure and energy densities. Another way to describe cosmological 
acceleration is a modification of the classical action of the general relativity (GR)  by an addition of a non-linear 
function of curvature, $F(R)$:
\be
A =-\frac{\mpl^2}{16\pi}\int d^4x\-g\,\left[R+F(R)\right] ,
\label{A}
\ee
where function $F(R)$ is chosen in such a way that the modified GR equations have a solution $R = const $ in
absence of any matter source. Taking the trace $g_{\mu\nu} \delta A/\delta g_{\mu\nu} =0$, we find:
\be\label{eq:trace}
3\D^2 F'_R-R+RF'_R-2F=  8\pi T^\mu_\mu/\mpl^2.
\ee
where $\D^2\equiv \D_\mu\D^\mu$ is the covariant D'Alambertian operator, $F'_R\equiv d F/ d R$, 
{and $T_{\mu\nu}$ is the energy-momentum tensor of matter.} 

To describe the astronomical data about cosmological acceleration the solution of the equation
\be
R_c-R_c F'_R (R_c) + 2 F(R_c) = 0
\label{eq-Rc}
\ee
should be equal to $R_c = - {32\pi \Omega_\lambda \rho_c}/{m_{Pl}^2}$, where $\Omega_\lambda \approx 0.75$ is the fraction
of the observed vacuum-like energy density and $\rho_c \approx 10^{-29}$ g/cm$^3$ is the total cosmological energy density.

The pioneering suggestion~\cite{one-over-R} of gravity modification with $F(R) \sim \mu^4/R$
suffered from strong instabilities in celestial bodies~\cite{DolgKaw}. To cure these instability 
further modifications of gravity have been suggested~\cite{HuSaw,ApplBatt,Starob}. 
These and some other versions are reviewed e.g. in refs.~\cite{rev-f-of-R,noj-odin}. 
The suggested modifications, however, may lead to infinite-$R$ singularities 
in the past cosmological history~\cite{rev-f-of-R} and in the future in astronomical systems
with rising energy/matter density~\cite{frolov, Arb_Dolgov}. 
These singularities can be successfully eliminated 
by an addition of the $R^2$-term into the action. Such a contribution naturally appears as a result of  
quantum corrections due to matter loops in curved space-time~\cite{Gur-Star,Starobinsky_1980}. Possibly in 
astrophysical systems singularity may be avoided even without $R^2$ if one takes into account distortion
of the background flat Minkowsky space-time by an impact of rising $R$. However, it would take place if $R$
very much differs from its GR value and such a deviation from normal gravity would be observable.

In what follows we consider the version of the modified gravity suggested by Starobinsky~\cite{Starob}:
\be\label{eq:model}
{ F(R) = -\lambda R_0\left[1-\left(1+\frac{R^2}{R_0^2}\right)^{-n}\right]-\frac{R^2}{6m^2},
}\ee
where $n$ is an integer, $\lambda>0$, and $| R_0 | \sim 1/t_U^2$,  where $t_U \approx 13$ Gyr is the universe age.
Parameter $m$ is bounded from below, $m\gtrsim 10^5$ GeV, to preserve 
successful predictions of  BBN, see e.g. \cite{Arb_Dolg_Rev}.

Other models\cite{HuSaw,ApplBatt} demonstrate similar behavior. 
We study the evolution of $R$ in a contracting astrophysical system and show that $R$ oscillates around its GR 
value with quite a large amplitude, efficiently producing elementary particles with masses below the oscillation 
frequency. The particle production damps the oscillations and may suppress or even eliminate the singularity, which 
would appear if $R^2$ term was absent.

Cosmology with $R^2$ term and in particular the particle production by the oscillating curvature was studied in the 
early works~\cite{Zeld-Star, Starobinsky_1980,Vilenkin_1985}. There was renewed interest to this 
problem~\cite{Arb_Dolg_Rev,japanese} stimulated by the study of the interplay of the infrared and ultraviolet
terms in eq. (\ref{eq:model}).

In this paper we study solutions of eq. (\ref{eq:trace}) in the system of non-relativistic particles with rising energy density 
which we approximate as $\rho = \rho_0 (1 + t/t_{contr}) $, where $t_{contr}$ is the effective time of contraction. We assume 
that the gravity of matter is not strong and thus the background metric can be considered as flat. 
Written in terms of $R$ equation of motion (\ref{eq:trace}) contains non-linear derivative terms:
\be
[(\partial_t R)^2 - (\partial_j R)^2 ]/R,
\label{non-linear}
\ee
which makes it very inconvenient for qualitative analysis. To avoid that, we introduce, instead of  $R$, the new 
function, proportional to $F'(R)$:
\be 
\xi\equiv  \frac{1}{2\lambda n}\left(\frac{T_0}{R_0}\right)^{2n+1}F'_R, 
\label{eq:xi}
\ee
where $ T_0 = 8\pi \rho_0/m_{Pl}^2 $. It is also convenient to introduce dimensionless quantities 
$ z = {\rho (t)}/{\rho_{0}}$,  $y = -{R}/{T_0}$,
$g =  1/ [ 6\lambda n  (m t_U)^2] \left({\rho_{m0}}/{\rho_c}\right)^{2n+2} $, and dimensionless time
$\tau = m t \sqrt g $. In terms of these quantities and in the limit of $R\gg R_0$ and sufficiently homogeneous
$\xi$ (both conditions are naturally fulfilled) the equation of motion for $\xi$ takes a very simple form:
\be\label{eq:xi_evol}
\xi''+ dU/d\xi =0\,,
\ee
where prime denotes derivative with respect to $\tau$ and ${dU}/{ d\xi}= z - y(\xi)$. Unfortunately $y$ cannot be expressed
through $\xi$ analytically. We have only inverse relation 
\be
\xi = {1}/{y^{2n+1}} -gy , 
\label{xi-of-y}
\ee
so we have to use different approximate 
expressions in different ranges of $\xi$.

In what follows we make analytical estimates of the amplitude of the curvature oscillations and compare them with numerical
calculations for some values of the parameters. The agreement is generally quite good. However, we cannot do numerical 
calculations for realistically tiny values of $g$ or huge frequencies of oscillations due to numerical instability and/or too long
computational time. So an agreement of the analytical results with numerical ones where the later can be done allows to conclude
that analytical estimates can be valid for high frequency and small $g$ as well.

The minimum of potential $U(\xi)$ is located at $y(\xi) = z(\tau)$, so it moves with time according to 
\be
\xi_{min} (\tau) ={z(\tau)^{-(2n+1)}}-gz(\tau). 
\label{xi-min}
\ee
It is intuitively clear that even if initially $\xi$ takes its GR value $\xi = \xi_{min}$ it would not catch the motion of the minimum
and as a result it starts to oscillate around it. Dimensionless frequency of small oscillations, $\Omega$, is determined by:
\be
\label{eq:frequency_U}
\Omega^2 =  \left. \frac{\partial^2 U}{\partial \xi^2}\right|_{y=z} = \left(\frac{2n+1}{z^{2n+2}}+g\right)^{-1}\,.
\ee
Note that physical frequency is $\omega = \Omega\,m \sqrt g$.

For small oscillations $\xi= \xi_{min} + \xi_1$ takes the form:
\be
\xi(\tau)\approx \xi_{min}(\tau)+\alpha(\tau)\sin [F(\tau)+\delta] 
\label{eq:xi_expansion}
\ee
where  $\delta$ is a phase determined by the initial conditions and
\be
F (\tau) \equiv \int^\tau_{\tau_0}d\tau'\,\Omega(\tau').
\label{F-of-tau}
\ee

For positive and negative $\xi$ the potential can be approximated as:
\be
\label{eq:xi_potential}
U(\xi) = U_+(\xi)\Theta(\xi) + U_-(\xi)\Theta(-\xi)\,,
\ee
where
\be
\label{eq:potential_+_-}
\begin{aligned}
U_+(\xi) &= z\xi - \frac{2n+1}{2n}\left[\left(\xi+g^{(2n+1)/(2n+2)}\right)^{2n/(2n+1)}-g^{2n/(2n+2)}\right]\,,\\
U_-(\xi) &= \left(z-g^{-1/(2n+2)}\right)\xi+\frac{\xi^2}{2g}\,.
\end{aligned}
\ee
There is a kind of the conservation law for the energy of  field $\xi$:
\be\label
{eq:conserved}
\frac{1}{2}\,\xi'^2 + U(\xi) - \int^\tau_{\tau_1}d\tau'\frac{\partial z}{\partial\tau'}\,\xi(\tau')=\text{ const},
\ee
where $\tau_1$ is an arbitrary fixed time moment.
The last term appears because $U$ explicitly depends on time through $z$. 
If ${ \partial z/\partial \tau}$ is positive, which is the case for a contracting body, the value of $U(\xi)$ 
{would} in general grow with time. According to assumption made above $z$ linearly grows with time
as $ z(\tau) = 1+\kappa\tau$, where 
\be
\kappa = (m t_{contr} \sqrt{g})^{-1}.
\label{kappa}
\ee
This simple law may be not accurate when $t/t_{contr}>1$, but probably the results obtained are not too far 
from realistic case.

Equation of motion \rif{eq:xi_evol} for small oscillations $\xi_1$ can be rewritten  as
\be
\label{eq:xi_evol_approx} {
\xi_1''+\Omega^2\xi_1= -\xi_{min}'' \,,}
\ee
{ Term} $\xi_{min}''$ is proportional to $\kappa^2$, which is usually 
assumed small, so in first approximation it can be neglected, though an analytic solution { for constant ${ \Omega}$ or in the limit of large ${ \Omega}$} 
can be obtained with this term as well. 
Using expansion \rif{eq:xi_expansion} and neglecting $\alpha''$, we obtain
\be
\label{eq:alpha_omega}
 \alpha\simeq \alpha_0\,
\sqrt\frac{{ \Omega_0}}{{ \Omega}}= \alpha_0\left(\frac{1}{z^{2n+2}}+\frac{g}{2n+1}\right)^{1/4}
{ \left({1}+\frac{g}{2n+1}\right)^{-1/4}}.
\ee
{Here and in what follows sub-0 means that the corresponding quantity is taken at initial moment $\tau = \tau_0$.}

{ We impose the following initial conditions}
\be
\begin{cases}
y(\tau=\tau_0)=z(\tau=\tau_0)=1\,,\\
y'(\tau=\tau_0)=y'_0\,,
\end{cases}
\label{init-y}
\ee
the first of which corresponds to GR solution at the initial moment. 
The initial value of the amplitude
derivative $\alpha_0$ can be expressed through $y_0'$, which we keep as a free parameter, as: 
\be 
{ \alpha_0 = (\kappa - y'_0) (g+ 2n+1)^{3/2} }.
\label{alpha-0}
\ee

\begin{figure}
\centering
\includegraphics[width=.4\textwidth]{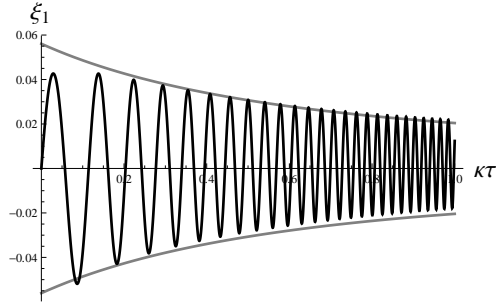}
\caption{Oscillations of $\xi_1(\tau)$, in the case $n=2$, ${ \kappa=0.01}$, ${ g=0.01}$ and initial conditions $y_0=1$, $y'_0=\kappa/2$. 
The amplitude of the oscillations is in very good agreement with analytical estimate \rif{eq:xi_expansion}.}
\label{fig:xi_amplitude}
\end{figure}

To keep $\xi_1$ small the initial value of the derivative $y'_0$ should be also small. In this case
the numerical results, shown in figure \ref{fig:xi_amplitude}, are in remarkable agreement with eq.~\rif{eq:xi_expansion}.
For $y' = \kappa$ the oscillations seem not to be excited. However, this is an artifact of the approximation used.
For example, the "source" term in the r.h.s. of eq.~(\ref{eq:xi_evol_approx}) creates oscillations and hence deviations from GR
with any initial conditions.

\begin{figure}[!t]
\centering
 \includegraphics[width=.4\textwidth]{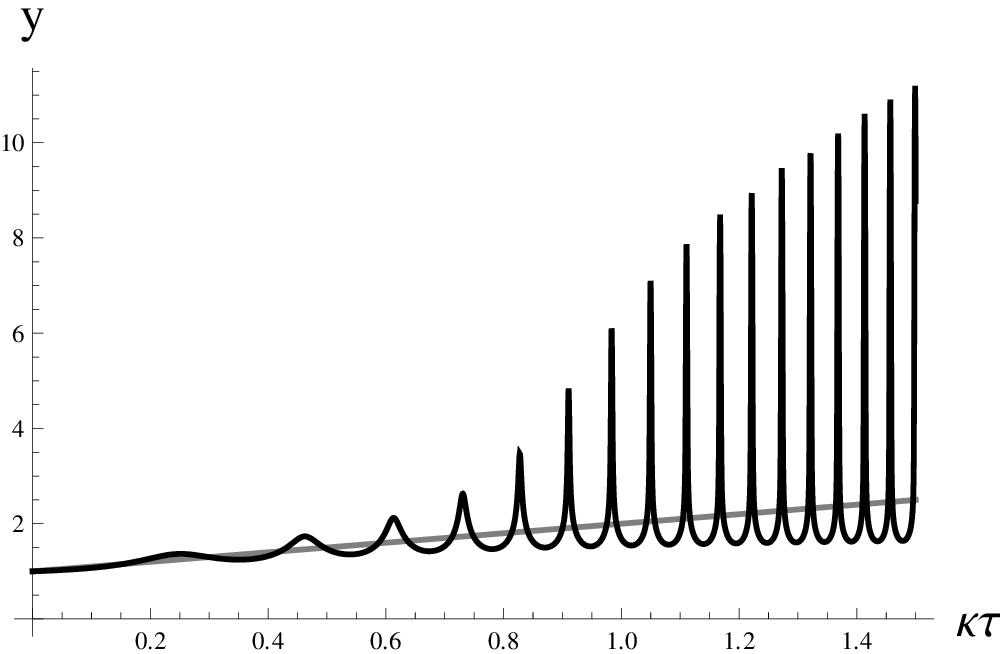}
\includegraphics[width=.4\textwidth]{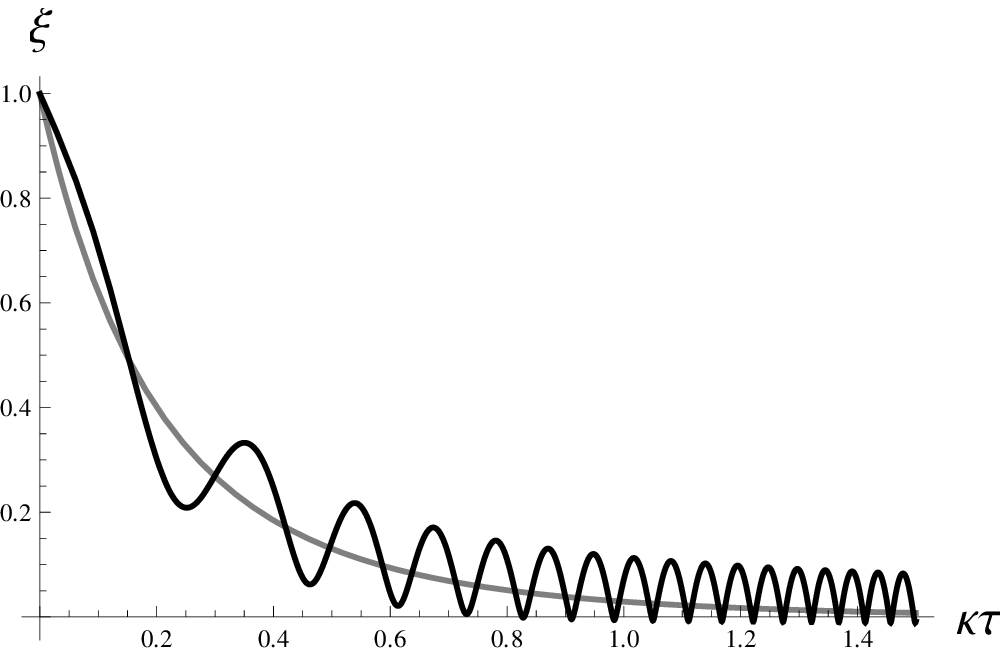}
\caption{``Spikes'' in the solutions. The results presented are for $n=2$, $g=0.001$, $\kappa=0.04$, and $y'_0=\kappa/2$. 
Note the asymmetry of the oscillations of $y$ around $y=z$ and their anharmonicity.}
\label{fig:spikes}
\end{figure}

Our primary goal is to determine the amplitude and shape of the oscillations of $y$. In contrast to  $\xi$ the oscillations of $y$ are 
strongly unharmonic and for negative and even very small $| \xi |$  the amplitude of $y$ may be very large because
$y \approx -\xi /g$, according to eq. (\ref{xi-of-y}). This feature is well demonstrated by the results of numerical calculations 
shown in figure \ref{fig:spikes}.

We can estimate the amplitude of the spikes using the energy evolution law (\ref{eq:conserved}).
Let us use it at the moment when the maximum value of $\xi$ reaches zero. So at this moment $\xi' = 0$. 
Since $U(0) =0$, the constant in the r.h.s. of eq. (\ref{eq:conserved}) turns to zero. Now let us go to larger time and neglect the 
oscillating part of $\xi$ under the integral. The maximum absolute value of negative $\xi$ is determined by the
equation:
\be
U_- (\xi_{max}) = \kappa \int_{\tau_1}^\tau d\tau'  \left[ z^{-(2n+1)} - gz\right] = 
\frac{1}{2n} \left[ \left(\frac{1}{z(\tau_1)}\right)^{2n} - \left(\frac{1}{z(\tau)}\right)^{2n} \right]
+\frac{1}{2}g \left[ {z^2(\tau_1) - z^2(\tau)} \right].
\label{U-of-xi-max}
\ee
The value of $z(\tau_1) \equiv z_1$ is found from the
condition $\alpha = \xi_{min}$ i.e.
\be 
z_1^{-(2n+1)} - g z_1 = |\kappa - y_0' | \left( g +2n +1\right)^{5/4}\,\left(\frac{2n+1}{z_1^{2n+2}} + g\right)^{1/4}
\label{z1-eq}
\ee
In the limit of small $g$, $g< 1/z_1^{2n+2}$,  asymptotically $|\xi_{max}| = (g/n)^{1/2}\, z(\tau_1)^{-n}$. In the same limit  eq. (\ref{z1-eq}) gives:
\be
z_1 = \left[ \left( \kappa - y_0' \right)^2\,\left(2n+1 \right)^3 \right]^{-1/(3n+1)}.
\label{z1}
\ee
This finally determines 
\be
y_{max} = (ng)^{-1/2} z_1^{-n}.
\label{y-max}
\ee
This result is much larger than the naive estimate 
$g y_{max} ^{2n+2} \sim 1$ \cite{Arb_Dolgov}.  For negative $\xi$ potential (\ref{eq:potential_+_-}) behaves as $U \approx \xi^2/(2g)$, so 
the characteristic frequency of the oscillations  in the region of negative $\xi$
is about $\Omega \sim 1/\sqrt{g}$ in dimensionless time which corresponds to $\omega \approx m$ in physical time.
Evidently frequency of oscillations of $y$ in this region is the same. It can be shown that the calculated amplitude 
of $y$ (\ref{y-max}) becomes noticeably larger with rising $z$~\cite{Arb_Dolg_Rev_preparation}.
 
According to calculations of  ref.~\cite{Arb_Dolg_Rev}, harmonic oscillations of curvature with frequency $\omega$
and amplitude $R_{max}$ transfer energy to massless particles with the rate (per unit time and volume):
\be
\label{eq:PP_harmonic}
\dot\rho_{PP}\simeq {R^2_{max}\,\omega}/{(1152\pi)}\,,
\ee
The life-time of such oscillations is $\tau_R = {48\,\mpl^2}/{\omega^3} $.

In our case the oscillations are far from harmonic and we have to make Fourier expansion of the spiky function $y(\tau)$.
To this end we approximate the ``spike-like'' solution as a sum of gaussians of amplitude modulated by slowly varying
amplitude ${ B(t)}$, superimposed  on smooth background $ A(t)$:
\be
\label{eq:gaussians}
R(t) = A(t)+B(t)\sum_{j=1}^{N}\exp\left[-\frac{(t-jt_1)^2}{2\sigma^2}\right]\,.
\ee
We  assume that $ \sigma\ll t_1 $, { that is} the spacing between spikes is much larger than their width. 
The Fourier transform of { expression} \rif{eq:gaussians} is straightforward but rather tedious (details can be
found in our work~\cite{Arb_Dolg_Rev_preparation}).
Finally we find:
\be
\left|\tilde{\mc R}(\omega)\right|^2\simeq \frac{4\pi^2 B^2
\sigma^2e^{-\omega^2\sigma^2}\Delta t}{t_1^2}\sum_j\delta\left(\omega-\frac{2\pi j}{t_1}\right)\,.
\ee
Identifying $B$ with $R_{max} = y_{max}  T_0$ and integrating over frequencies we obtain
\be
 \dot\rho_{PP} \simeq 
{y_{max}^2 T_0^2}/({576\pi\,t_1})\,.
\label{dot-rho-total}
\ee
Time interval $t_1$ is approximately equal to $2\pi/\omega_{slow}= 2\pi/(\Omega_{slow}m \sqrt{g})$, 
where $\Omega_{slow}$ is given by eq.~\rif{eq:frequency_U}. Amplitude $y_{max}$ is defined in eq.~(\ref{y-max});
parameter $\kappa$ is determined by eq. (\ref{kappa}) and "coupling constant" $g$ is defined below eq. (\ref{eq:xi}).
Taking all the factors together we finally obtain:
\be
\dot \rho_{PP} = C_n \frac{\rho_0^2 (m t_U)^2\, z^{n+1}}{m_{Pl}^4 t_{contr}}\left( \frac{t_U}{t_{contr} }\right)^{\frac{n-1}{3n+1}}\,
\left( \frac{\rho_c}{\rho_0} \right)^{\frac{(n+1)(7n+1)}{3n +1}},
\label{dot-rho-fin}
\ee
where
\be
C_n = (2n+1)^{\frac{9n-1}{2(3n+1)}} \left( 6\lambda n \right)^{\frac{7n+1}{2(3n +1)}}/ (18 n).
\label{c-n}
\ee

It is convenient to present numerical values: $\rho_c/ m_{Pl}^4 \approx 2\cdot 10^{-123}$ and $ (mt_U)^2 \approx 3.6\cdot 10^{93} m_5^2$, where
$\rho_c \approx 10^{-29}$g/cm$^3$ and  $m_5 = m/10^5$ GeV. Now assuming that the particle production lasts during $t \approx t_{contr}$
and taking $\rho_0 = \rho_c$, we find the energy flux of cosmic rays produced by oscillating curvature:
\be
\rho_{CR} \approx 10^{-24} m_5^2 \, z^{n+1} \, {\rm GeV\, s^{-1}\, cm^{-2}}.
\label{rho-CR}
\ee
This result is a lower limit of the flux of the produced particles. With larger $z$ when the minimum of the potential 
shifts deep into the negative $\xi$ region the production probability significantly rises. Analytical evaluation in this case is 
more difficult but simple qualitative arguments indicate to it and numerical calculations supports this assertion.

Recall that our derivation has been done under assumption $R>R_0$ and respectively $\rho_0 > \rho_c$. However if we take $\rho_0$ somewhat larger than $\rho_c$ the results would be approximately correct.    

For $m = 10^{10}$ GeV the predicted flux of cosmic rays with energy around $10^{19}$ eV is close or even higher than the 
observed flux.

\subsection*{Acknowledgement} 
EA and AD acknowledge the support of the
grant of the Russian Federation government 11.G34.31.0047.


\begin{thebibliography}{99}



\bibitem{one-over-R}
S. Capozziello, S. Carloni, A. Troisi,
\textit{Recent Res. Dev. Astron. Astrophys.} {\bf 1} (2003) 625,
arXiv:astro-ph/0303041;\\
S.M. Carroll, V. Duvvuri, M. Trodden, M.S. Turner,
\textit{Phys. Rev.} {\bf D 70} (2004) 043528, arXiv:astro-ph/0306438.

 \bibitem{DolgKaw}
A.D. Dolgov, M. Kawasaki, \textit{Phys. Lett.} {\bf B 573} (2003) 1.

 \bibitem{HuSaw}
 W. Hu, I. Sawicki, \textit{Phys. Rev. D} {\bf 76}, 064004 (2007).
 
 \bibitem{ApplBatt}
 A. Appleby, R. Battye, \textit{Phys. Lett. B} {\bf 654}, 7 (2007).

\bibitem{Starob}
A.A. Starobinsky, \textit{JETP Lett.} {\bf 86}, 157 (2007).

\bibitem{rev-f-of-R}
S.A. Appleby, R.A. Battye, A.A. Starobinsky, JCAP 1006 (2010) 005.

\bibitem{noj-odin}
S. Nojiri, S. Odintsov, Phys.Rept. 505 (2011) 59;\\
K. Bamba, S. Capozziello, S. Nojiri. S.D. Odintsov, arXiv: 1205.3421.

\bibitem{frolov}
A.V.  Frolov,  \textit{Phys. Rev. Lett.} {\bf 101}, 061103 (2008);\\
I. Thongkool, M. Sami, R. Gannouji, S. Jhingan, \textit{Phys. Rev. D} {\bf 80} 043523 (2009);\\
I. Thongkool, M. Sami, S. Rai Choudhury, \textit{Phys. Rev. D} {\bf 80} 127501 (2009).
 
 \bibitem{Arb_Dolgov}
E.V. Arbuzova, A.D. Dolgov, \textit{Phys. Lett.} {\bf B} 700, 289 (2011).


 \bibitem{Gur-Star} 
 V.Ts. Gurovich, A.A. Starobinsky, 
      \textit{Sov. Phys. JETP} \textbf{50} (1979) 844; [Zh. Eksp. Teor. Fiz. 77 (1979) 1683];\\
      A.A. Starobinsky, \textit{JETP Lett.} \textbf{30} (1979) 682; 
      [Pisma Zh. Eksp. Teor. Fiz. \textbf{30} (1979) 719];\\
             A.A. Starobinsky, ``Nonsingular model of the Universe with the 
   quantum-gravitational de Sitter stage and its observational consequences'',
      in \textit{Proc. of the Second Seminar "Quantum Theory of Gravity"} (Moscow,
      13-15 Oct. 1981), INR Press, Moscow , 1982, pp. 58-72; reprinted in:
      Quantum Gravity, eds. M. A. Markov and P. C. West. Plenum Publ. Co.,
      N.Y., 1984, pp. 103-128.

 \bibitem{Starobinsky_1980}
A.A. Starobinsky, \textit{Phys. Lett.} \textbf{B91}, 99 (1980).

 \bibitem{Zeld-Star}
Ya. B. Zeldovich, A.A. Starobinsky, \textit{JETP Lett.} \textbf{26} (1977) 252. 

\bibitem{Vilenkin_1985} A. Vilenkin,      \textit{Phys. Rev.} \textbf{D32}, 2511 (1985).

\bibitem{Arb_Dolg_Rev}
E.V. Arbuzova, A.D. Dolgov, L. Reverberi, \i{JCAP} \b{02}, 049 (2012).

\bibitem{japanese}
H. Motohashi, A. Nishizawa, arXiv:1204.1472 [astro-ph.CO].

 \bibitem{Arb_Dolg_Rev_preparation}
E.V. Arbuzova, A.D. Dolgov and L. Reverberi,  in preparation.


%
%
%
%
%
%
%
%
%



%
%

%
%
%
%

%
%
%
%
%
%




\end{thebibliography}
\end{document}